\documentclass[aps,prb]{revtex4}
\usepackage{amsmath,amssymb}
\usepackage{graphicx}
\usepackage{dcolumn}
\usepackage{bm}
\usepackage{color}
\usepackage{relsize}
\newcommand{\mbb}{\mathbb}
\newcommand{\mc}{\mathcal}

\newcommand{\tet}{\texttt}

\begin{document}
\title{Exploring stable long-lifetime plasmon excitations in the Lieb lattice}
\author{
Andrii Iurov$^{1}$\footnote{E-mail contact: aiurov@mec.cuny.edu, theorist.physics@gmail.com},
Liubov Zhemchuzhna$^{1,2,3}$\footnote{E-mail contact: lzhemchuzhna@mec.cuny.edu, lzhemchuzhna@fordham.edu},
Godfrey Gumbs$^{3}$, 
and
Danhong Huang$^{4}$}

\affiliation{
$^{1}$Department of Physics and Computer Science, Medgar Evers College of City University of New York, Brooklyn, NY 11225, USA\\ 
$^{2}$Department of Physics \& Engineering Physics, Fordham University, Bronx, NY 10458, USA\\
$^{3}$Department of Physics and Astronomy, Hunter College of the City University of New York, 695 Park Avenue, New York, New York 10065, USA\\ 
$^{4}$Space Vehicles Directorate, US Air Force Research Laboratory, Kirtland Air Force Base, New Mexico 87117, USA
}
\date{\today}

\begin{abstract}
The subject of the present paper is a thorough numerical investigation of plasmon expectations, their dispersions and damping within a Lieb lattice. The Lieb lattice is known for its unique low-energy band structure which consists of a bandgap as well as a flat band intersecting the conduction band at its lowest point. In contrast to previously studied dice lattice, the location of  the current flat band exhibits reduced and broken symmetries, which give rise to interesting electronic and optical properties of this new material. In this work, we have investigated the conditions for observing a well-defined and stable plasmon mode within a wide frequency range. Specifically, we have considered a free-standing layer with various doping levels, as well as different types of monolayers of the Lieb lattice interacting with a surface-plasmon mode localized on top of a semi-infinite conductor. In particular, we have observed and described fully long-living plasmon modes  with unusual energy dispersions. Additionally, we have carried out a detailed investigation on the static screening associated with the Lieb lattice. Our study has further revealed that these predicted features seem to be quite different from those of pseudospin-1 materials but resemble those of graphene instead. \end{abstract}
\maketitle

\section{Introduction} 
\label{sec1}

All  recently discovered naturally occurring and fabricated two-dimensional  (2D) materials have received tremendous attention by researchers in various fields due to graphene with its unusual electronic properties.\,\cite{geim2007rise, geim2009graphene}  Included among these are included  materials with spin-orbit coupling\,\cite{galitski2013spin, mawrie2014magnetotransport} in which an electron’s spin interacts with a self-induced magnetic field by its own orbital motion and leads to the spin-split energy bands as well as strong spin Hall effect\,\cite{sinova2015spin, jungwirth2012spin}. Meanwhile, tremendous progress has been made investigating twisted bilayer graphene\,\cite{hou2024strain, naumis2021reduction, navarro20233} exhibiting new and promising electronic phenomena due to hybridized Dirac cones since the Brillouin zone in this case becomes much smaller than that of a regular monolayer graphene.
Among these well-known 2D lattices, we would like to highlight materials with a flat (dispersionless) band in their low-energy band structure. One of the best known and well-studded models for such materials is the so-called $\alpha$-$\mc{T}_3$ model, which is represented by a 2D lattice which interpolates between graphene and the dice lattice. Its unit-cell contains three atoms (sites), {\em i.e.\/} two rim sites $A$ and $B$, which make up a hexagon, and a hub atom $C$  locating at the center of each hexagon. Consequently, a zero-energy flat band is found for all types of the $\alpha$-$\mc{T}_3$ materials. Here, $\alpha$ is the relative hopping parameter between the hub-hub and rim-hub electron transitions, indicating a major difference from the graphene model having $\alpha = 0$. The other limiting case for $\alpha=0$ is known as a dice lattice.

\medskip
\par
Various physical properties, including optical,\,\cite{tamang2023probing} magnetic,\,\cite{raoux2014dia,zhang2024quantum,islam2023role,balassis2020magnetoplasmons} electronic,\,\cite{weekes2021generalized,tamang2023orbital,iurov2023application,gorbar2019electron, islam2023properties,iurov2019peculiar} collective,\,\cite{oriekhov2020rkky,roldan2011theory,malcolm2016frequency} thermal, transport\,\cite{oriekho2023quantum,iurov2020klein,illes2017klein,iurov2020quantum} and topological\,\cite{vidal2001disorder,vidal1998aharonov} behaviors, have been investigated extensively in recent times. Importantly, there has been a series of very interesting and promising works on how the electron band structure could be altered by applying off-resonance dressing fields with various polarizations. These include ones applied to graphene,\,\cite{mojarro2020dynamical,kibis2010metal, kristinsson2016control} other lattices with Dirac cone\,\cite{iurov2022floquet,ibarra2019dynamical} and the flat-band materials\,\cite{dey2018photoinduced,iurov2024floquet}. A present-day search for two-dimensional lattices with a flat band resulted in an interesting finding of moire superlattices formed by twisted bilayers of  two-dimensional materials, which is a novel, unusual and a very interesting class of quantum materials.\,\cite{pan2025lithography} The flat band in these materials lead to the existence of to heavy electrons which demonstrate ceratin similarity to magic-angle twisted bilayer graphene.

\par
An important example of the flat band lattices is the Lieb lattice. Its atomic composition is based on a square lattice  in which one of the sites is removed. Consequently, one is left with a band structure with a finite gap between the valence and conduction bands, in addition to a flat band intersecting with the conduction band at its lowest point, {\em i.e.\/} located right above the band gap.\,\cite{kajiwara2016observation, mukherjee2015observation, slot2017experimental,nictua2013spectral} Another crucial representative of the flat band family of materials is a kagome lattice,\,\cite{jo2012ultracold,guo2009topological,mojarro2023topological,lee2024atomically} {\em i.e.\/} a triangle-based lattice forming a hexagonal pattern. Kagome lattices, which have been realized experimentally,\,\cite{wang2024dispersion} often host non-trivial topology and Chern insulators\,\cite{guo2009topological, xue2019acoustic} as well.

\medskip
\par

Plasmons, which refer to the collective charge density oscillations of nearly all conducting electrons within a metallic lattice, have become one of the most important phenomena in condensed-matter physics and many-body physics. A crucial part of studying plasmons is often connected to low-dimensional materials. Plasmonic phenomena have been studied extensively in a 2D electron gas,\,\cite{agarwal2014long,backes1992dispersion,badalyan2009anisotropic, jalali2018tilt} graphene,\,\cite{politano2014plasmon,yan2012infrared,pyatkovskiy2008dynamical, wunsch2006dynamical, polini2008plasmons, sarma2013intrinsic, hwang2007dielectric} buckled honeycomb lattices,\,\cite{tabert2014dynamical} transition metal dichalcogenides,\,\cite{sriram2020hybridizing, andersen2013plasmons, PhysRevB.88.035135} Kekule-distorted graphene,\,\cite{herrera2020electronic,andrade2019valley} flat band materials, such as dice lattice, $\alpha$-$\mc{T}_3$ model,\,\cite{malcolm2016frequency, iurov2022finite} as well as anisotropic and tilted Dirac cones\,\cite{milicevic2019type,wild2023optical,gomes2021tilted,mojarro2021optical, sadhukhan2017anisotropic, yan2022anomalous}. It also includes several recently discovered unusual lattice systems\,\cite{dutta2022collective,yerin2023dielectric, dey2022dynamical, dutta2023intrinsic, hayn2021plasmons, sadhukhan2020novel, torbatian2021hyperbolic, stauber2013optical} with Rashba spin-orbit coupling\,\cite{wang2005plasmon,shitrit2013spin}. In addition, plasmons have been discovered and studied in graphene and $\alpha$-$\mc{T}_3$ based nanoribbons,\,\cite{fei2015edge, iurov2021tailoring, brey2007elementary,karimi2017plasmons,roslyak2010unimpeded} exotic spherical graphitic particles and fullerenes\,\cite{gumbs2014strongly, ju1993excitation}, heterostructures \,\cite{sarma1981collective,li2017first,yao2018broadband} and system which includes two-dimensional layers coulomb coupled with conducting surfaces\,\cite{gumbs2015nonlocal}. The key experimental technique for investigating collective electronic excitations is the Electron energy loss spectroscopy (EELS). The dielectric response, plasmon energies (frequencies), as well as the momentum transfer of electron transitions, strengths of coupling to external light could be measured using angle-resolved energy loss spectroscopy, photoemission spectroscopy and ellipsometry. Both long- and short-wavelength excitations could be scanned by these techniques.

\par
In addition, theoretical studies of plasmon are covered by calculating the dynamical polarization function which is used for several important calculations and models, such as static screening, Boltzmann conductivity\,\cite{trescher2015quantum} and optical conductivity\,\cite{oriekhov2022optical, iurov2023optical,wareham2023optical,tan2021anisotropic,xiong2023optical}
for nearly all known two-dimensional materials.

\medskip
\par

Various sections of plasmonics, and especially surface plasmons and plasmon-polariton excitations,\,\cite{zhang2012surface, berini2012surface, barnes2006surface} demonstrate tremendous potential for various technological applications.\,\cite{koseki2016giant, petrov2017amplified,simon1983inhomogeneous}
There has also been a vast amount of experimental work done on plasmons in low-dimensional materials.\,\cite{garcia2014graphene, constant2016all, garcia2014graphene}
In our previous studies\,\cite{zhemchuzhna2024polarizability}, we have predicted that the plasmon mode of a Lieb lattice does not exist when its doping level is equal to the bandgap, and the numerical solution for the dielectric function could only approach zero within a very narrow region of the energy and wave-vector, although it never reaches zero exactly. Therefore, the crucial goal of this work is to find the conditions under which we can observe a well-defined  and long-living plasmon mode.  
\medskip
\par

The remaining part of our paper is organized as follows. We first present the model in 
Sec.\,\ref{sec2} for a low-energy Hamiltonian, band structure and corresponding electronic states (wave functions and eigenenergies) of the Lieb lattice with a flat band.  In Sec.\,\ref {sec3}, we discuss mathematical formalism and corresponding numerical results for dynamical polarization function and plasmon excitation in Lieb Lattices. For Sec.\,\ref{sec4}, we calculate and analyze the plasmon excitations in two Coulomb-coupled layers of the Lieb lattice. Our numerical results and discussions are followed in Sec.\,\ref{sec5} by considering the plasmon excitations in a system of one Lieb-lattice layer interacting with a semi-infinite conductor. 
A detailed comparison is shown in Sec.\,\ref{sec6} for the dynamical polarization function and plasmon excitations in a Lieb and dice lattice under similar conditions, {\em e.g.\/} electron doping. Section\ \ref{sec7} deals with the static screening of a charged impurity within a Lieb lattice in comparison with other cases having a doped atom in different flat-band materials. Finally, our conclusions, summary statements and 
remarks are presented in Sec.\,\ref {sec8}.

\medskip
\medskip
\par
\section{Electronic states in a Lieb lattice}
\label{sec2}

We begin with the basic form for the Hamiltonian, energy dispersions and the electronic eigenstates for the Lieb lattice. The low-energy Hamiltonian can be written as\,\cite{oriekhov2022optical}

\begin{equation}
\label{lieb106} 
\mc{H}^{(\text{\,L})}(\mbox{\boldmath$k$}\, \vert \, k_\Delta) = \hbar v_F \, \left\{
\begin{array}{c c c}
 k_\Delta & k_x & 0 \\[0.1cm]
 k_x & - k_\Delta  & k_y \\[0.1cm]
 0 & k_y & k_\Delta 
\end{array}
\right\} \ , 
\end{equation}
where the following translation of the wave vector
\begin{equation}
k_{x,y} \rightarrow \frac{\pi}{a_0} + k_{x,y}
\label{subst}
\end{equation}
is assumed to express the Hamiltonian in Eq.\,\eqref{lieb106} starting from its general form. In Eq.\,\eqref{subst} and $a_0$ is the lattice parameter.

\medskip 

The Hamiltonian given in Eq.\,\eqref{lieb106} yields three solutions for the energy subbands 

\begin{eqnarray}
\varepsilon^{(\text{\,L})}_{\gamma = \pm 1} (\mbox{\boldmath$k$}\, \vert \, k_\Delta)/\hbar v_F & = & \gamma \sqrt{ k_\Delta^2 + k_x^2 + k_y^2} \ , \\
\nonumber 
\varepsilon^{(\text{\,L})}_{\gamma = 0} (\mbox{\boldmath$k$}\, \vert \, k_\Delta)/\hbar v_F & = & k_\Delta \ ,
\end{eqnarray}
which could be simplified as one unified expression, given by 

\begin{equation}
\varepsilon^{(\text{L})}_{\gamma}(\mbox{\boldmath$k$}\, \vert \, k_\Delta) = \hbar v_F \, \left[\delta_{\gamma, 0} \, k_\Delta + \gamma \,(1 - \delta_{\gamma, 0})\, \sqrt{k_\Delta^2 + k^2} \,\right] \ , 
\label{lieb116}
\end{equation}
where $\gamma=0,\,\pm 1$, and $\delta_{\gamma, 0}$ is the Kronecker delta. From Eq.\,\eqref{lieb116}, we find that one of the solutions with $\gamma=0$ represnts a flat band located at $\varepsilon  = \hbar v_F k_\Delta$. In contrast, the bandgap in a dice lattice is described by a $\hat{\Sigma_z^{(3)}}$-term, given by 

\begin{equation}
\label{gapdice01}
\delta \mc{H}^{(D)}(\Delta_0) = \Delta_0 \, \hat{\Sigma_z^{(3)}} = \Delta_0  \, \left\{
\begin{array}{ccc}
1 & 0 & 0 \\
0 & 0 & 0 \\
0 & 0 & -1
\end{array}
\right\} \ . 
\end{equation}
Consequently, the energy spectrum of a dice lattice will be simply given by a flat band $\varepsilon_{\gamma = 0}(k) = 0$, which is always located at the zero energy level, as well as symmetric valence and conduction bands $\varepsilon_{\gamma = \pm 1}(k)  = \gamma \, \hbar v_F \sqrt{k^2 + \Delta_0^2}$. 
\medskip 

Correspondingly, the wave functions are obtained as  

\begin{eqnarray}
\label{liebwg}
{\bf \Psi}^{(\text{L})}_{\gamma = \pm 1} (\mbox{\boldmath$k$}\, \vert \, k_\Delta) & = & 
\frac{1}{\sqrt{2 E_k (E_k - \gamma\hbar v_Fk_\Delta)}}
\,
\left\{
\begin{array}{c}
k_x \\
- k_\Delta + \gamma E_k \\
k_y
\end{array}
\right\}\ ,
\end{eqnarray}
and 

\begin{equation}
\label{liebw0}
{\bf \Psi} ^{(\text{L})}_{\gamma = 0} (\mbox{\boldmath$k$}\, \vert \, k_\Delta) =   \frac{1}{k} \, \left\{
\begin{array}{c}
- k_y \\
0\\
k_x
\end{array}
\right\} = \left\{
\begin{array}{c}
- \sin \Theta_{\bf k} \\
0\\
\cos \Theta_{\bf k}
\end{array}
\right\}\ ,
\end{equation}
where $E_k\equiv\hbar v_F\,\sqrt{k_\Delta^2+k^2}$ and $\Delta_0=2\hbar v_F  k_\Delta$ is the bandgap similarly to the badgap in a dice lattice. 
\medskip

The wave functions given by Eqs.\,\eqref{liebwg} and \eqref{liebw0} can be used to compute the overlaps $\mathcal{O}_{\gamma, \gamma'} (\mbox{\boldmath$k$}, \mbox{\boldmath$q$}\,\vert\,k_\Delta)\equiv\Big| \Big \langle {\bf \Psi} ^{(\text{L})}_{\gamma} (\mbox{\boldmath$k$}+ \mbox{\boldmath$q$} \, \vert \, k_\Delta) \, \Big \vert \, {\bf \Psi} ^{(\text{L})}_{\gamma'} (\mbox{\boldmath$k$}\, \vert \, k_\Delta)
\Big \rangle  \Big|^2 $ factor. A straight-forward but lengthy calculation leads to the following results 

\begin{eqnarray}
\label{mainOgL1}
&& \mathcal{O}_{\gamma =  \pm 1, \gamma' =  \pm 1 } (\mbox{\boldmath$k$}, \mbox{\boldmath$q$}\,\vert\,k_\Delta) = \frac{ \left\{ 
k^2 +  \mbox{\boldmath$k$} \cdot \mbox{\boldmath$q$}  +  \left(
E_{{\bf k} + {\bf q}}  - \gamma k_\Delta
\right) \, \left(
E_k - \gamma k_\Delta
\right)
\right\}^2
}{2 E_k \, E_{{\bf k} + {\bf q}} \, \left(
E_k - \gamma k_\Delta
\right) \left(
E_{{\bf k} + {\bf q}} - \gamma' k_\Delta
\right) } \ , 
\end{eqnarray}
and 

\begin{eqnarray}
\label{mainOgL2}
&& \mathcal{O}_{\gamma = \pm 1, \gamma' = 0}(\mbox{\boldmath$k$}, \mbox{\boldmath$q$}\,\vert\,k_\Delta) =  
\frac{
k^2 \left|\mbox{\boldmath$k$} + \mbox{\boldmath$q$} \right|^2 - \left[ k^2 + (\mbox{\boldmath$k$} \cdot \mbox{\boldmath$q$}) \right]^2
}{2 E_k \left(
E_k - \gamma k_\Delta \right)
\left|\mbox{\boldmath$k$} + \mbox{\boldmath$q$} \right|^2 }
\ . 
\end{eqnarray}
\medskip

\begin{figure} 
\centering
\includegraphics[width=0.45\textwidth]{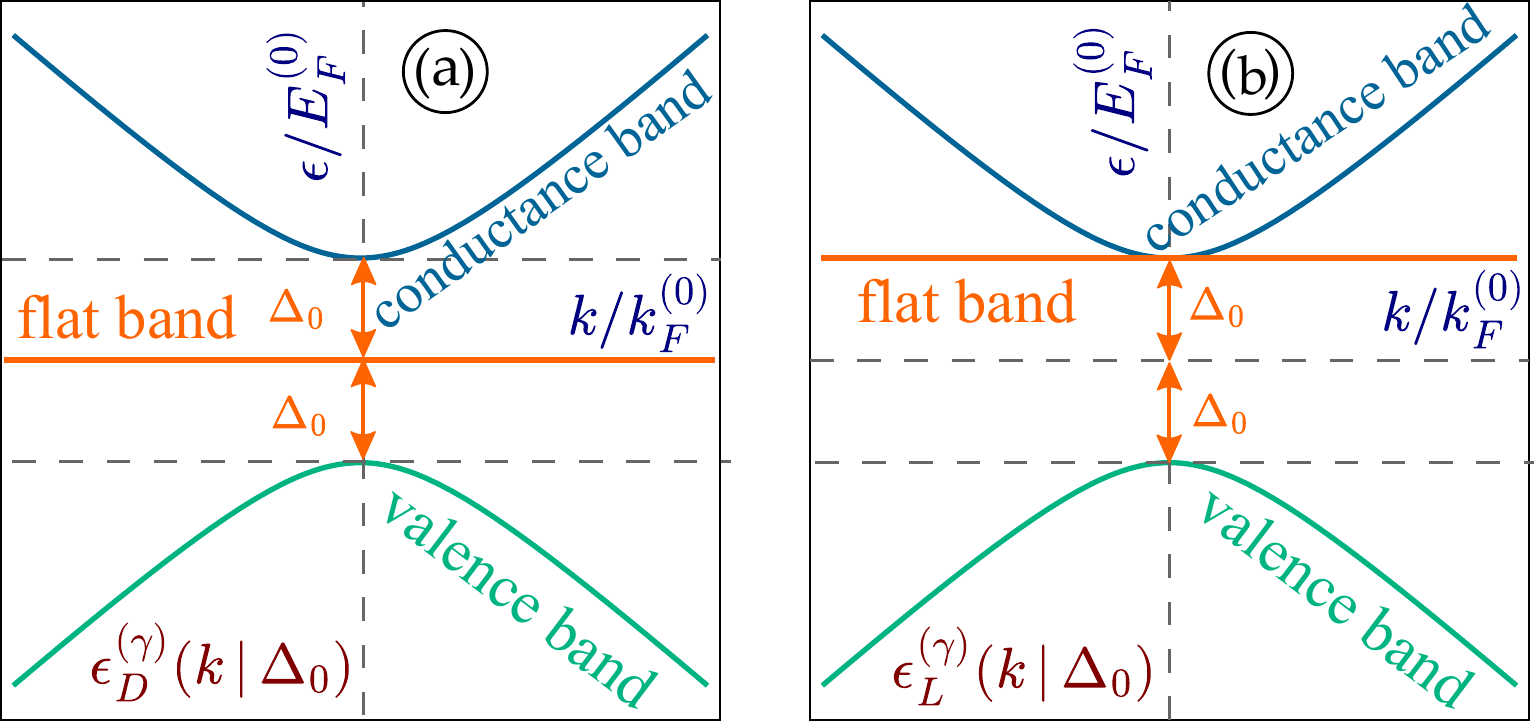}
\caption{(Color online) The low-energy band structures (the energy dispersions) of a gapped dice lattice (left panel) and the Lieb lattice (right panel). These two dispersions demonstrate a lot of similarities, {\em e.g.\/} the presence of a flat band between the valence and conduction bands. However,  the locations of these two flat bands are quite different. For a Lieb lattice, the flat band intersects with the conduction band at its lowest point, which is in contrast with a completely symmetric case for a dice lattice.}
\label{FIG:1}
\end{figure}
\medskip

In Fig.\,\ref{FIG:1}, we present  calculated low-energy electron band structures obtained from Eq.\,\eqref{lieb116} for a dice lattice (left panel) as well as for a Lieb lattice (right panel). By comparing directly these two panels, we observe that these two types of band structure are quite similar to each other, which are obtained from the same  $3 \times 3$ pseudospin-1 Hamiltonians. Here, both lattices exhibit symmetrically located valence and conduction bands with a general type of dispersion $\backsim \sqrt{k^2 + \Delta_0^2}$,  bandgap equal to $2 \Delta_0$,  and flat (dispersionless) bands between the valence and conduction bands.  However, the location of this flat band in these two cases is very different. For a Lieb lattice, it intersects with the conduction band at its lowest point. In contrast, for a dice lattice,   the location of this flat band can be found exactly in the middle between the valence and conduction bands. Therefore, the specific band structure of the Lieb lattice becomes much less symmetric and leads to symmetry breaking related to new topological properties of electrons in this material. Such unique energy dispersions also imply rich new physics for electron transitions among these three bands, which becomes a key ingredient in calculating the dielectric function employed to determine the plasmon spectrum of this material.

\medskip
\par

Here, we would like to emphasize again that the key focus of this study deals with an attempt to understand how this elevated and less symmetric position of the flat band affect plasma dispersions and other major collective properties of the Lieb lattice in comparison with previously investigated dice lattice, as well as the $\alpha$-$\mc{T}_3$ model. 
\medskip

\begin{figure} 
\centering
\includegraphics[width=0.75\textwidth]{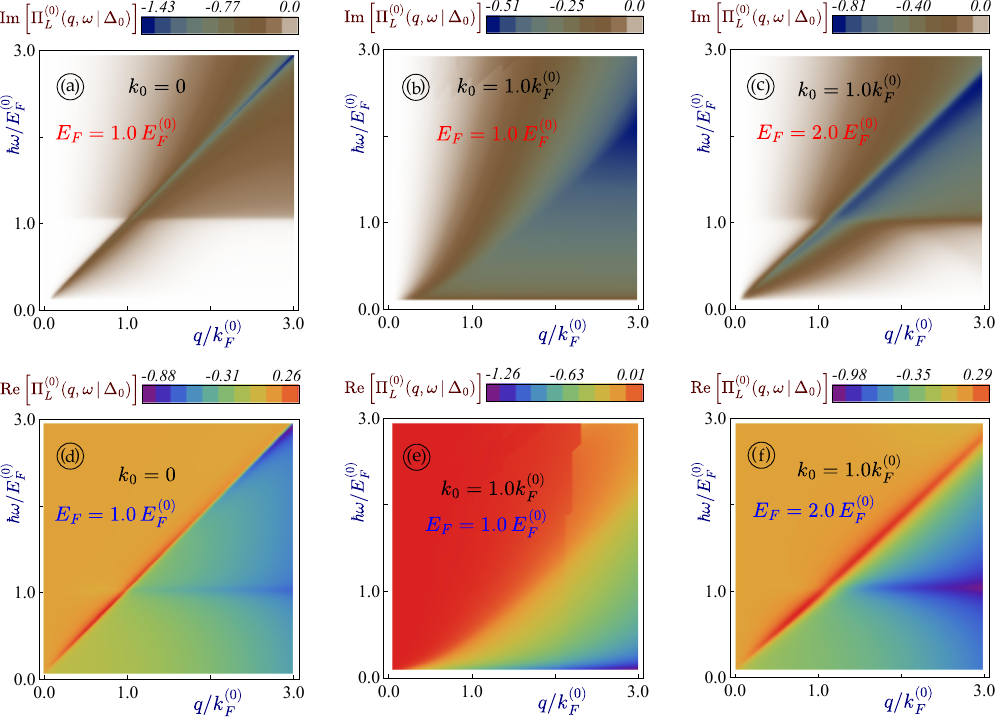}
\caption{(Color online) Numerically calculated dynamical polarization function $\Pi^{(0)}(q,\omega \, \vert \, \Delta_0)$ for a Lieb lattice. Here, each of upper panels  $(a)$, $(b)$ and $(c)$ displays the imaginary part of $\Pi^{(0)}(q,\omega \, \vert \, \Delta_0)$ and the particle-hole modes (or the single-particle excitation regions), corresponding to a finite imaginary part of the polarization function. The three lower panels $(d)$, $(e)$ and $(f)$ represent the real part of $\Pi^{(0)}(q,\omega \, \vert \, \Delta_0)$ as functions of both wave vector $q$  and frequency $\omega$. Two left panels $(a)$ and $(d)$ connect to a zero-gap model of a Lieb lattice with $\Delta_0 = 0$, and the energy dispersions are equivalent to those of a zero-gap $\alpha$-$T_3$  model. The other four panels correspond to an actual Lieb lattice with $k_0 = 1.0\, k_F^{(0)}$ for different Fermi energies (doping levels) $E_F=1.0\,E_F^{(0)}$ and $E_F=2.0\,E_F^{(0)}$, as labeled.}
\label{FIG:2}
\end{figure}

\begin{figure} 
\centering
\includegraphics[width=0.55\textwidth]{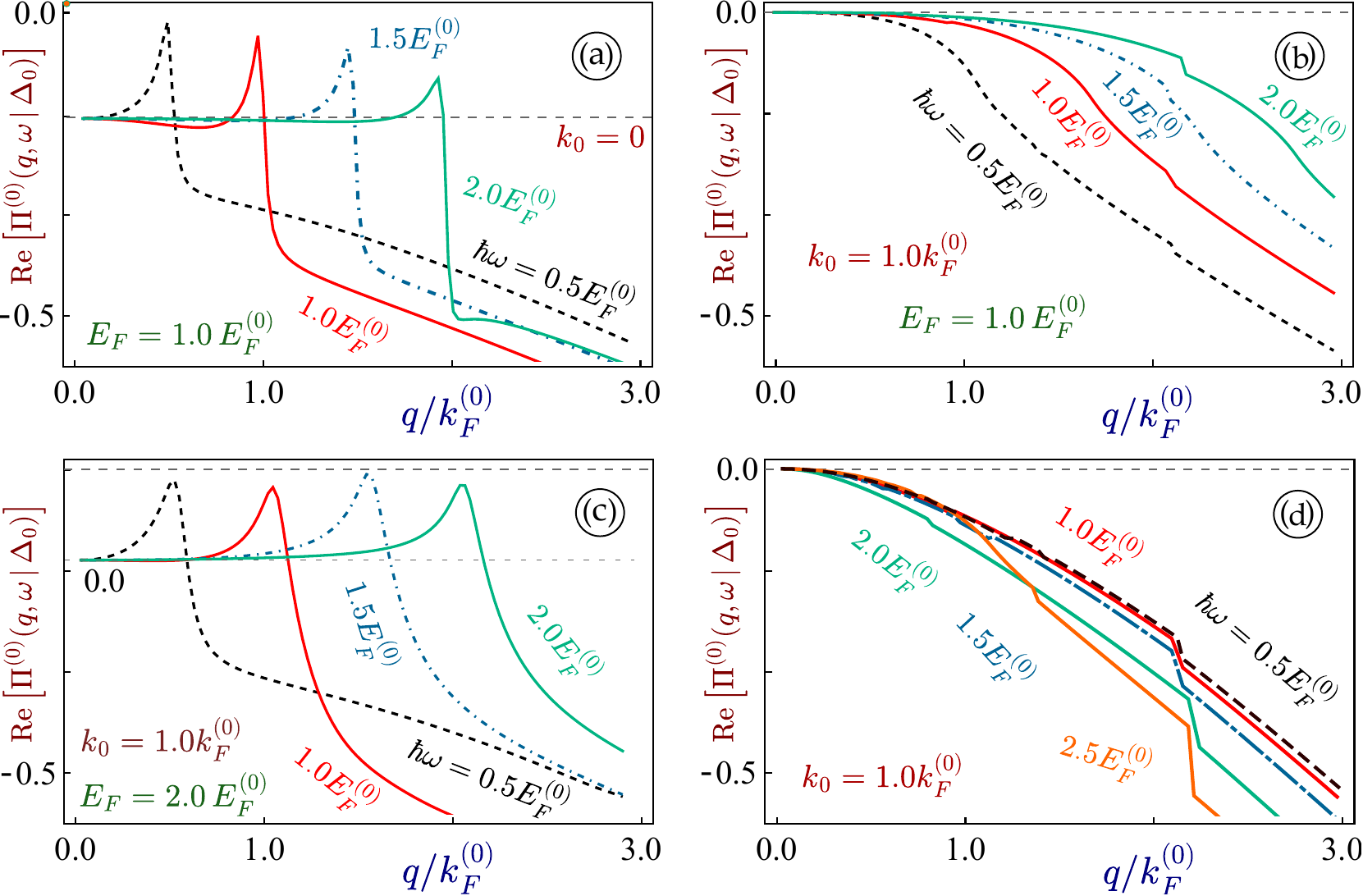}
\caption{(Color online) Constant frequency cuts to the real part of the polarization function $\Pi^{(0)}(q,\omega \, \vert \, \Delta_0)$ for a Lieb lattice as a function of wave vector $q$. Panel $(a)$ and $(c)$ deal with a zero-gap model for a Lieb lattice, corresponding to  $k_0 = 0$, and the energy dispersions become equivalent to those of a zero-gap $\alpha$-$T_3$ model. All the other plots  $(b)$, $(c)$ and $(d)$ are associated with an actual Lieb lattice with $k_0 = 1.0\, k_F^{(0)}$ with different Fermi energies (doping levels) $E_F=0.9\,E_F^{(0)}$ and $E_F=2.0\,E_F^{(0)}$ and $E_F=1.0\,E_F^{(0)}$, respectively. Each curve connects to a specific fixed frequency $\omega = 0.5\,E_F^{(0)}/\hbar$, $1.0\,E_F^{(0)}/\hbar$, $1.5\,E_F^{(0)}/\hbar$, $2.0\,E_F^{(0)}/\hbar$ and $2.5\,E_F^{(0)}/\hbar$ for all panels, as labeled.}
\label{FIG:3}
\end{figure}

\begin{figure} 
\centering
\includegraphics[width=0.49\textwidth]{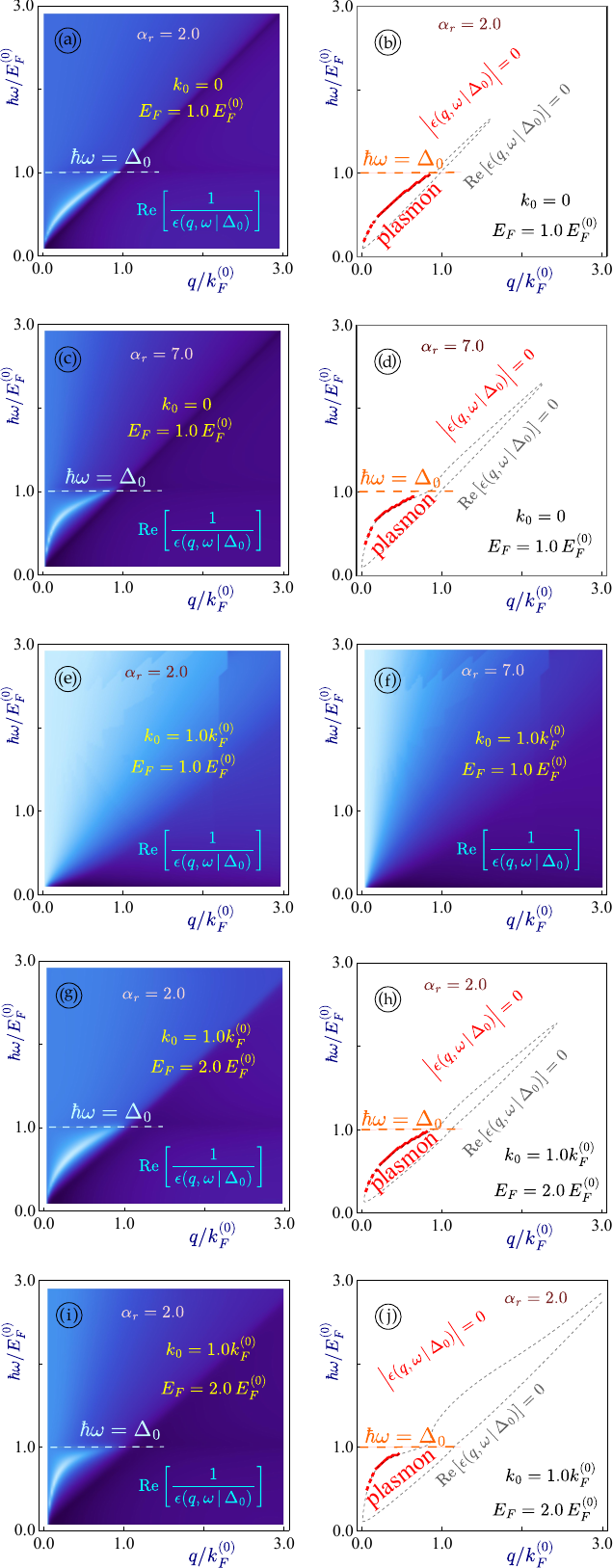}
\caption{(Color online) Plasmon dispersions for a monolayer Lieb lattice. The left panels $(a)$,  $(c)$, $(e)$,  $(g)$ and $(i)$, as well as the right panel $(f)$, represent density plots of the real part of inverse dielectric function $1/\epsilon(q,\omega \, \vert \, \Delta_0)$ whose peaks correspond to the plasmon modes. The broadened peaks of $1/\epsilon(q,\omega \, \vert \, \Delta_0)$, on the other hand, reveal damped plasmons. Here, the right-hand-side panels $(b)$,  $(d)$, $(h)$ and $(j)$ display the exact numerical solutions for the plasmon branches. The Landau damped plasmons are presented in gray dashed curves while undamped ones are given by red solid curves. The first two upper rows (plots $(a)$ - $(c)$) connect to a zero-gap model with $k_0=0$ for a Lieb lattice,  and then, the energy dispersions are equivalent to those of a zero-gap $\alpha$-$T_3$  model. The remaining panels $(d)$ - $(h)$ correspond to a regular Lieb lattice with the same $k_0 = 1.0 \,k_F^{(0)}$ but different Fermi energies (doping levels) $E_F=1.0\,E_F^{(0)}$ and $E_F=2.0\,E_F^{(0)}$. Meanwhile, we also present the plasmon branches for different values of the relative dielectric constant $\alpha_r=2.0$ and $7.0$, as labeled.}
\label{FIG:4}
\end{figure}

\begin{figure} 
\centering
\includegraphics[width=0.75\textwidth]{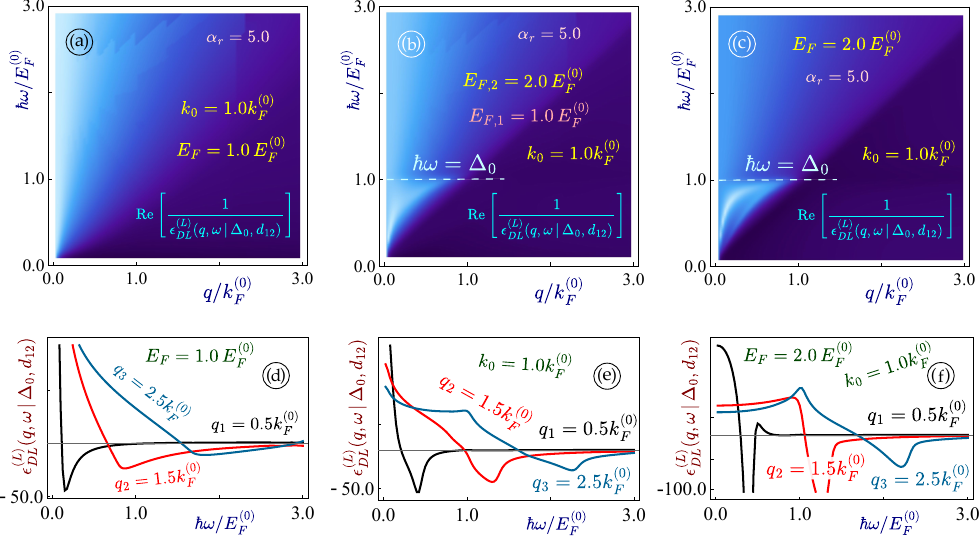}
\caption{(Color online) Plasmon dispersions for two Lieb lattice Coulomb-coupled monolayers. Here, three top panels $(a)$,  $(b)$ and $(c)$ represent the density plots for the real part of an inverse effective dielectric function $1/\epsilon_{DL}^{(L)}(q,\omega \, \vert \, \Delta_0)$ for a Coulomb-coupled double layer whose peaks represent the plasmon modes. The broadened peaks of $1/\epsilon^{(L)}(q,\omega \, \vert \, \Delta_0)$, on the other hand, display damped plasmons. The lower panels $(b)$,  $(d)$ and $(e)$ present the frequency dependence of the constant wave vector cuts of inverse effective dielectric function $1/\epsilon_{DL}^{(L)}(q,\omega \, \vert \, \Delta_0)$. The left panes $(a)$ and $(c)$ connect to the case of two identical Lieb lattices with $k_0 = 1.0 k_F^{(0)}$. The middle panels $(b)$ - $(e)$ are associated with two regular Lieb lattice with $k_0 = 1.0 \,k_F^{(0)}$ and different Fermi energies (doping levels) $E_F=1.0\,E_F^{(0)}$ and $E_F=2.0\,E_F^{(0)}$, and two equivalent Lieb lattices with $E_F=2.0\,E_F^{(0)}$ are presented in panels $(c)$ and $(f)$. All results for plasma branches correspond to the relative dielectric constant $\alpha_r=5.0$, as labeled. }
\label{FIG:5}
\end{figure}

\begin{figure} 
\centering
\includegraphics[width=0.55\textwidth]{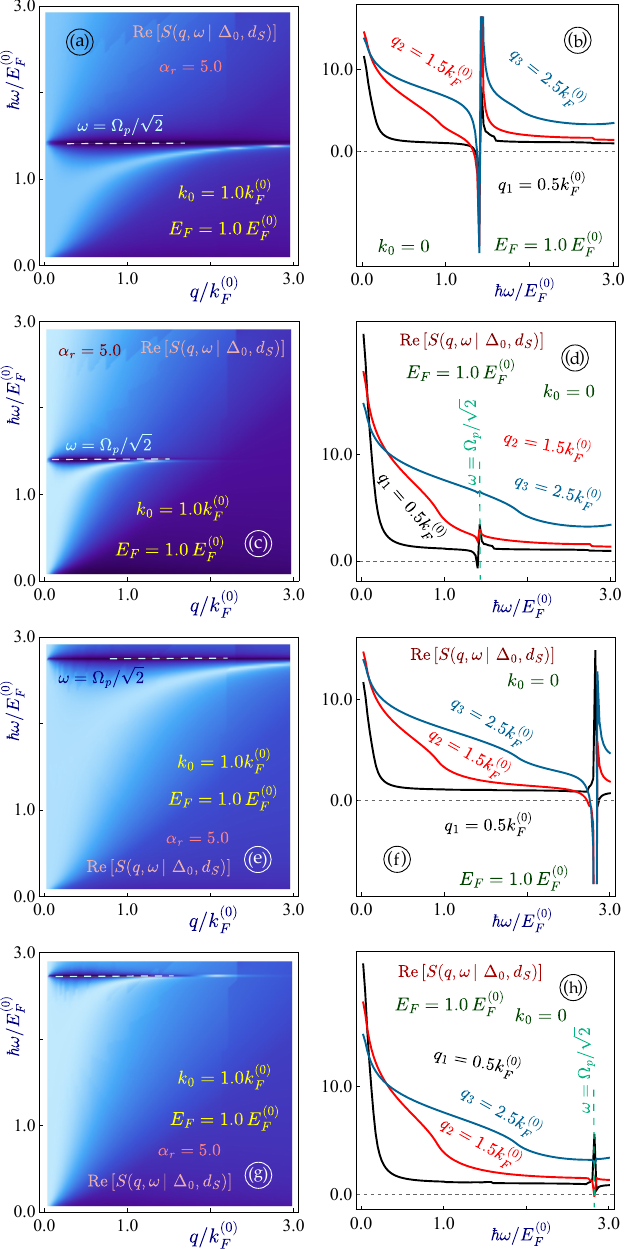}
\caption{(Color online)  
Plasmon dispersions for a Lieb lattice monolayer Coulomb-coupled to semi-infinite conductor. The left-hand-side panels $(a)$,  $(c)$ and $(e)$ represent the density plots of the inverse dispersion function $\mathcal{S}^{-1}(q,\omega \, \vert \, \Delta_0,d_s)$ whose peaks correspond to the plasmon modes. The broadened peaks of $\mathcal{S}^{-1}(q,\omega \, \vert \, \Delta_0,d_s)$ reveal damped plasmons. The right-hand-side panels $(a)$,  $(c)$ and $(e)$ show the exact numerical solutions for the plasmon branches. The Landau damped plasmons are presented in dashed gray curves and undamped ones  - in red solid lines. In all  cases, the Fermi energy (doping level) is selected as $E_F=1.0E_F^{(0)}$ which corresponds to the location  of the flat band.  The distances between the layer and the conducting surface are chosen as $d=0.5 k_F^{(0)\,\, -1}$ and $1.5 k_F^{(0)\,\, -1}$, correspondingly, while the bulk plasmon frequencies in the metal are $\Omega_p = 2.0 E_F^{(0)}/\hbar$ and $4.0 E_F^{(0)}/\hbar$. All plots are provided for extrinsic (doped) Lieb lattices with the energy bandgap $\Delta_0 = 1.0\, E_F^{(0)}$.
}
\label{FIG:6}
\end{figure}

\begin{figure} 
\centering
\includegraphics[width=0.55\textwidth]{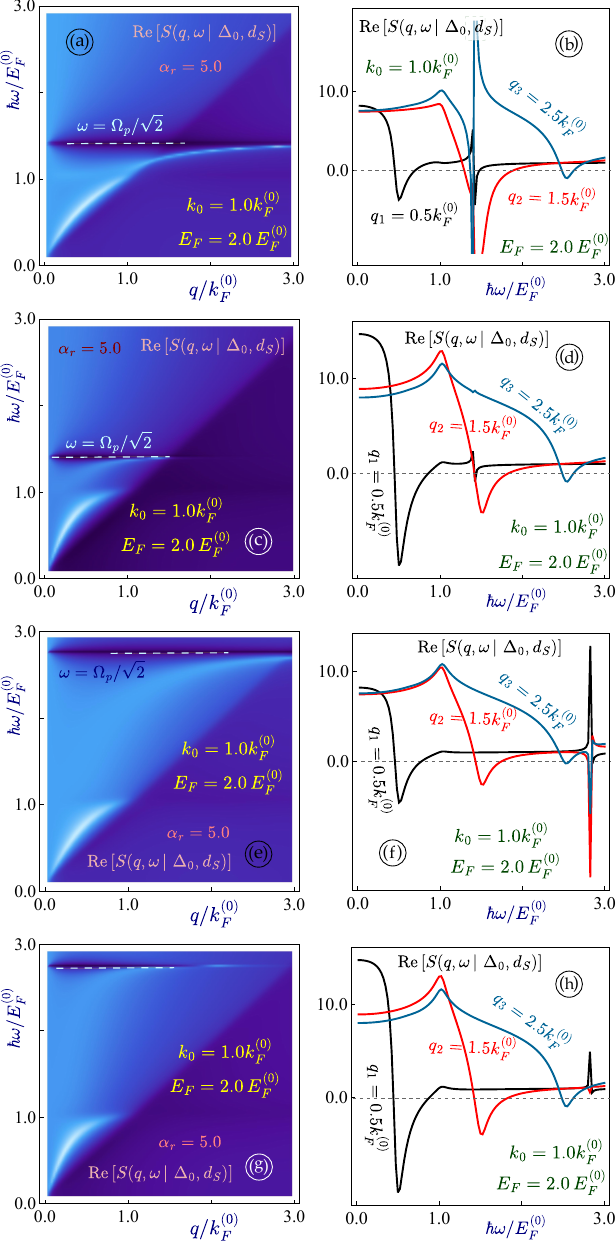}
\caption{(Color online) Plasmon dispersions for a Lieb lattice monolayer Coulomb-coupled to semi-infinite conductor. The left-hand-side panels $(a)$,  $(c)$ and $(e)$ represent the density plots of the inverse dispersion function $\mathcal{S}^{-1}(q,\omega \, \vert \, \Delta_0,d_s)$ whose peaks correspond to the plasmon modes. The broadened peaks of $\mathcal{S}^{-1}(q,\omega \, \vert \, \Delta_0,d_s)$ reveal damped plasmons. The right-hand-side panels $(a)$,  $(c)$ and $(e)$ show the exact numerical solutions for the plasmon branches. The Landau damped plasmons are presented in dashed gray curves and undamped ones  - in red solid lines. In all  cases, the Fermi energy (doping level) is selected as $E_F=2.0E_F^{(0)}$ which corresponds to a location in the conduction band (above the flat band). The distances between the layer and the conducting surface are chosen as $d=0.5 k_F^{(0)\,\, -1}$ and $1.5 k_F^{(0)\,\, -1}$, correspondingly, while the bulk plasmon frequencies in the metal are $\Omega_p = 2.0 E_F^{(0)}/\hbar$ and $4.0 E_F^{(0)}/\hbar$. All plots are provided for extrinsic (doped) Lieb lattices with the energy bandgap $\Delta_0 = 1.0\, E_F^{(0)}$.
}
\label{FIG:7}
\end{figure}

\begin{figure} 
\centering
\includegraphics[width=0.8\textwidth]{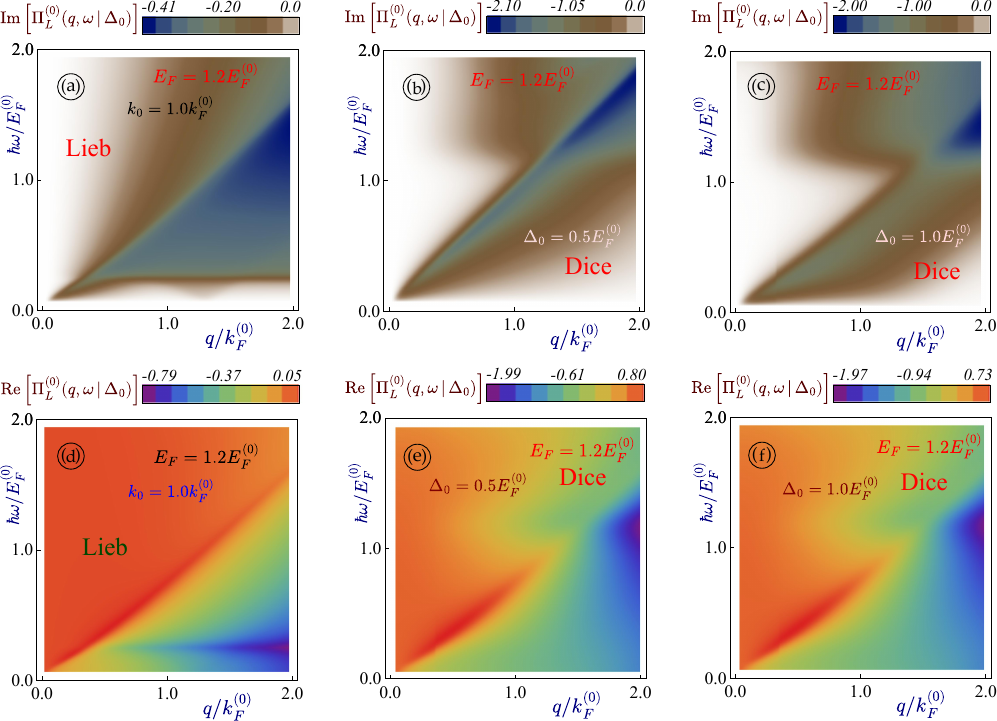}
\caption{(Color online) Dynamical polarization function $\Pi^{(0)}(q,\omega \, \vert \, \Delta_0)$ for a Lieb lattice and a dice lattice. We aim to compare the real and imaginary parts of $\Pi^{(0)}(q,\omega \, \vert \, \Delta_0)$, we well as the plasmon dispersions for the two likewise lattices with a different location of the flat band in similar conditions. The Fermi energies (doping levels) have been chosen $E_F=1.2E_F^{(0)}$  for all cases.  Each of the upper panels  $(a)$, $(b)$ and $(c)$ demonstrates the imaginary part of $\Pi^{(0)}(q,\omega \, \vert \, \Delta_0)$ and the particle home modes (or the single-particle excitation) regions, corresponding to a finite imaginary part of the polarization function.  The  three lower panels $(d)$, $(e)$ and $(f)$ represent the real part of $\Pi^{(0)}(q,\omega \, \vert \, \Delta_0)$ as a function of wave vector $q$  and frequency $\omega$. The two left panes $(a)$ and $(d)$ correspond to a Lieb lattice with $k_0 = 1.0 k_F^{(0)}$, while the remaining plot describe a dice lattice with different bandgaps $\Delta_0= 0.5\,E_F^{(0)}$ (plots $(b)$ and $(e)$) and $\Delta_0= 1.0\,E_F^{(0)}$ (panels $(c)$ and $(f)$), as labeled. }
\label{FIG:8}
\end{figure}

\begin{figure} 
\centering
\includegraphics[width=0.55\textwidth]{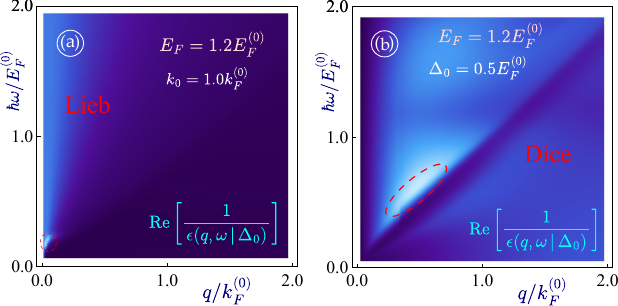}
\caption{(Color online) Plasmons for the Lieb lattice (panel $(a)$) and a dice lattice (panel $(b)$). We aim to compare the plasmon dispersions for the two likewise lattices with a different location of the flat band in similar conditions. The Fermi energies (doping levels) have been chosen $E_F=1.2E_F^{(0)}$  for all cases. All panes represent the density plots of the real part of the inverse dielectric function $\epsilon^{-1}(q,\omega \, \vert \, \Delta_0)$ whose peaks correspond to the plasmon modes. The broadened peaks of $\epsilon^{-1}(q,\omega \, \vert \, \Delta_0)$ reveal damped plasmons.}
\label{FIG:9}
\end{figure}

\begin{figure} 
\centering
\includegraphics[width=0.6\textwidth]{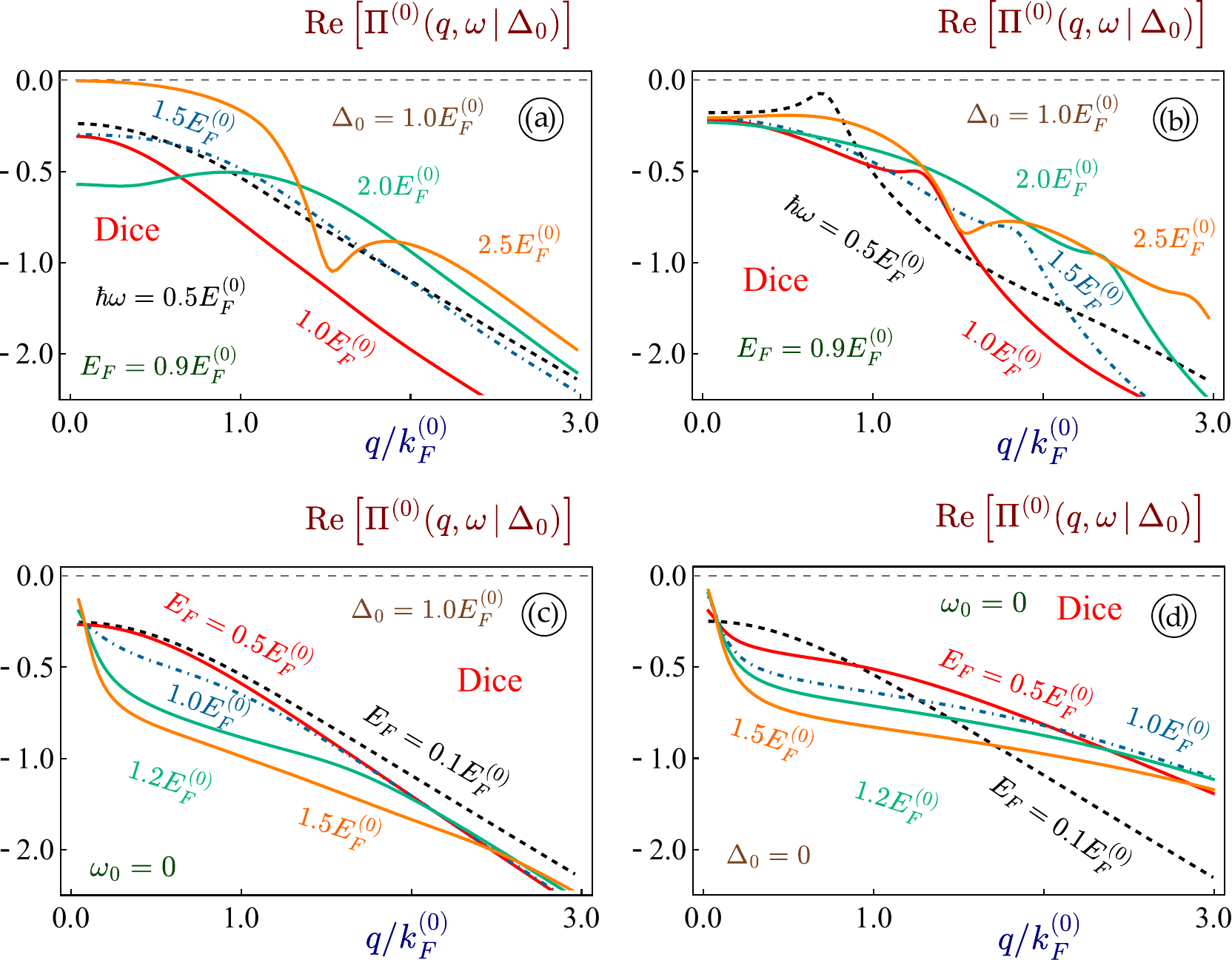}
\caption{(Color online) Constant frequency cuts of the real part of the polarization function $\Pi^{(0)}(q,\omega \, \vert \, \Delta_0)$ for a dice lattice as a function of wave vector $q$. In plots $(a)$ and $(b)$, each curve corresponds to a fixed Fermi energy (doping level), and in panels $(c)$ and $(d)$ - to a constant frequency ($\omega = 0.5\hbar/E_F^{(0)}$, $1.0\hbar/E_F^{(0)}$, $1.5\hbar/E_F^{(0)}$, $2.0\hbar/E_F^{(0)}$ and $2.5\hbar/E_F^{(0)}$ ), according to the provided labels. Specifically, panel $(c)$ demonstrates the static polarization with $\omega = 0$. }
\label{FIG:10}
\end{figure}

\section{Polarization function and plasmon dispersions for Lieb lattice}
\label{sec3}

Based on the electronic states of the Lieb lattice in Sec.\,\ref{sec2}, we now turn our attention to determining the plasmon modes in gapped dice and Lieb lattices From a theoretical point of view, the plasmon spectrum  of a 2D structure can be determined from the zeros of a dielectric function, given by

\begin{equation}
\epsilon(q, \omega\, \vert \, \Delta_0) =  1 - V_C(q) \, \Pi^{(0)}(q, \omega\, \vert \, \Delta_0) = 0 \ ,
\label{eps01}
\end{equation}
where $q$ and $\omega$ refer to the wave number and frequency, respectively,  of the plasmon excitation. In Eq.\,\eqref{eps01}, the Coulomb potential of interacting electrons within a 2D  material with background dielectric constant $\epsilon_r $ is written as $V_C(q)=2 \pi \alpha_r/q = e^2 /(2 \epsilon_0\epsilon_r \, q)$ with $\alpha_r = e^2/4\pi\epsilon_0 \epsilon_r$. From the solutions of Eq.\,\eqref{eps01}, we are able to obtain the dependence of frequency $\omega$ on the wave vector $q$, {\em i.e.\/}, a dispersion relation for the plasmon mode frequency $\omega=\Omega_{\rm pl}(q)$. Specifically, the dielectric function in Eq.\,\eqref{eps01} is given in terms of the polarization function $\Pi^{(0)}(q, \omega\, \vert \, \Delta_0)$ for interacting electrons.
Therefore, the essential quantity required for solving Eq.\,\eqref{eps01} is undoubtedly  the dynamical polarization function $\Pi^{(0)}(q, \omega\, \vert \, \Delta_0)$. In the lowest order for an external perturbation, the dynamical polarization function can be written explicitly as

\begin{equation}
\Pi^{(0)}(q, \omega \, \vert \, \phi, \mu(T)) = \frac{g}{4 \pi^2} \, \int d^2\mbox{\boldmath$k$} \sum\limits_{\gamma,\gamma' = 0 \, \pm 1} \, 
\mathfrak{O}_{\gamma, \gamma'} (\mbox{\boldmath$k$},\mbox{\boldmath$q$}\,  \vert\, k_\Delta) \,
\frac{n_F[\epsilon_\gamma(k\,\vert\,\Delta_0),\,\mu] - n_F[\epsilon_{\gamma'}(\vert \mbox{\boldmath$k$}+\mbox{\boldmath$q$}\vert\,\vert\,\Delta_0),\,\mu]}{ \left(\hbar \omega + i 0^+ \right) + \epsilon_\gamma  (k\,\vert\,\Delta_0) - \epsilon_{\gamma'} (\vert \mbox{\boldmath$k$} +\mbox{\boldmath$q$}\vert\,\vert\,\Delta_0)} \ .
\label{Pi00}
\end{equation}
In this notation, $n_F[\epsilon_\gamma (k\,\vert\, \Delta_0), \mu(T)] = \left\{1 + \tet{exp}[(\epsilon_\gamma (k\,\vert\,\Delta_0) - \mu)/(k_B T)] \right\}^{-1}$ in Eq.\,\eqref{Pi00} represents the thermal-equilibrium Fermi-Dirac distribution function. Moreover, the wave-function overlaps $\mathfrak{O}_{\gamma, \gamma'} (\mbox{\boldmath$k$}, \mbox{\boldmath$q$}\,  \vert\, k_\Delta)$ have already been computed in Eqs.\,\eqref{mainOgL1} and \eqref{mainOgL2}. 
\medskip

In our numerical computations, we have scaled energies, momenta and lengths by the Fermi momentum $k_F^{(0)}$ for pristine zero-gap graphene. Specifically, 
we set $k_F^{(0)} =  10^5 \, $cm$^{-1}$ corresponding to the electron density $n_0\backsim 10^{9}\,$cm$^{-2}$. Thus, the unit of energy is given by  
$E_F^{(0)} = \hbar v_F k_F^{(0)} \backsim 1$ or $2\,$meV, and the length scale is measured in units of $l = k_F^{(0)\,\,-1} = 100 \,$nm.
\medskip

Our numerical results for the dynamical polarization function $\Pi^{(0)}(q,\omega \, \vert \, \Delta_0)$, which becomes the essential building block for calculating the dielectric function $\epsilon(q,\omega \, \vert \, \Delta_0)$ of the Lieb lattice, are presented in Fig.\,\ref{FIG:2}. The two left panels of Fig.\,\ref{FIG:2} correspond to a ``toy model'' for a zero-bandgap Lieb lattice having $k_0=0$. In this case, the bandgap is closed, and then the valence and conduction bands form a regular Dirac cone. Moreover,  the flat band is symmetrically located and intersects both the valence and conduction bands at the Dirac point in this case. Consequently, the band structure becomes equivalent to the gapless $\alpha$-$\mc{T}_3$ model, or a dice lattice as a limiting case. However, the wave functions associated with their individual electronic states, as well as the resulting wave-function overlaps, remain different for thees two materials. Therefore, we can only claim that our current model reflects partially the zero-bandgap dice lattice, as a limiting case for the Lieb lattice having $k_0 \rightarrow 0$.  Meanwhile, we expect that both real and imaginary parts of $\Pi^{(0)}(q,\omega \, \vert \, \Delta_0)$ should share a similarity with known results of the $\alpha$-$\mc{T}_3$ model. 

\medskip
\par
We present the imaginary part of the dynamical polarization function in the upper panels of  Fig.\,\ref{FIG:2}. The regions with  $\text{Im} \left[ \Pi^{(0)}(q,\omega \, \vert \, \Delta_0) \right] \neq 0$ indicate the particle-hole mode or single-particle excitation regions, in which a plasmon is Landau damped  and a long-living plasmon cannot be sustained. Therefore, we are seeking regions with zero or insignificant values for  $\text{Im} \left[ \Pi^{(0)}(q,\omega \, \vert \, \Delta_0) \right]$ so that a self-sustained   plasmon mode could survive. First, we analyze the case with $E_F = 1.0 \, E_F^{(0)} = \Delta_0$, which  is related to  the doping level at (or just below) the flat band in the absence of electron transitions from the flat band to the conduction band. For this case, we have found that the plasmon is not well-formed, and regions free from the particle-hole modes do not exist. Such a situation changes dramatically for higher doping. As the Fermi  energy is located at $E_F = 2.0 \, E_F^{(0)} = \Delta_0$,
we observe a strong non-decaying plasmon mode and two nearby regions without single-particle excitation, {\em i.e.\/} a standard triangle satisfying $\hbar \omega < E_F$ and $ \omega > v_F q$ (above the diagonal), which was also seen in a dice lattice,\,\cite{malcolm2016frequency}, as well as additional regions (the bright areas in Fig.\,\ref{FIG:2} $(c)$) for $\hbar \omega < E_F$ and a large wave vector $q \geq 2 k_F$. This feature has not been observed in earlier studies of flat-band materials. Therefore, we conclude that a new type of plasmon and a novel structure of the particle's hole mode have been revealed for highly-doped Lieb lattices. 

\medskip
\par
The dependence  of several calculated results for chosen wave vector $q$ and  constant frequency plots of the dynamical polarization function $\Pi^{(0)}(q,\omega \, \vert \, \Delta_0)$ are presented in Fig.\,\ref{FIG:3}. Once again,  we separate the case with low doping $E_F = 1.0\,\Delta_0$ (or even $E_F = 0.9\,\Delta_0$ shown in panel $(d)$) and the case with higher doping $E_F = 2.0\,\Delta_0$, which allows for substantial intra-band electron transitions within the conduction band.  For the case when $E_F = 2.0\,\Delta_0$,  it is clear from the strong positive peaks in $\text{Re} \left[ \Pi^{(0)}(q,\omega \, \vert \, \Delta_0) \right]$ (also displayed by blue color in density plots of Fig.\,\ref{FIG:2}), which are missing in the case of $E_F=1.0\,\Delta_0$. In the case of $E_F = 2.0\,\Delta_0$, $\text{Re} \left[ \Pi^{(0)}(q,\omega \, \vert \, \Delta_0) \right]$ changes its sign, leading to zeros in dielectric function $\epsilon(q,\omega \, \vert \, \Delta_0)$. However, one does find a significant drop of all constant-frequency cuts of $\text{Re} \left[ \Pi^{(0)}(q,\omega \, \vert \, \Delta_0) \right]$ as $q \geq 1.5 k_F$, as displayed in panels $(b)$ and $(d)$ of Fig.\,\ref{FIG:3}. Obviously, these nearly monotonic and constantly negative values of $\text{Re} \left[ \Pi^{(0)}(q,\omega \, \vert \, \Delta_0) \right]$ cannot support the existence of a plasmon mode. 

\medskip 
\par
We now turn to the plasmon excitation in a free standing  layer of the Lieb lattice material, which is one of the principal subjects of this work. Our numerical results are presented in Fig.\,\ref{FIG:4}. Similar to our previous consideration of the polarization function, we discuss three separate cases: \  a ``toy model'' with $k_0=0$ for zero bandgap and the band structure identical to the gapless $\alpha$-$\mc{T}_3$ model (top panels), a regular Lieb lattice with under-critical doping ($E_F = \Delta_0$), as well as a regular Lieb lattice with higher doping $E_F =2.0 \,\Delta_0$. Apart from this, we further calculate the exact numerical solutions for the long-lifetime plasmon ({\em i.e.\/} the absolute value of the dielectric function is zero, or close to zero within certain precision in numerical evaluation), as well as a plasmon mode only with a limited lifetime, obtained only from zero of the real part of a  dielectric function given by Eq.\,\eqref{eps01}. 

\par 
Interestingly, the high-temperature behavior of flat-band Dirac materials is qualitatively similar to graphene. Specifically, the imaginary part of the polarization function falls off at  $\backsim 1/T$, while the real part of the polarization function increases with $T$.\,\cite{iurov2022finite} Therefore, we observe well-formed plasmons with a long lifetime at large temperatures. The reason for that with the increasing temperature, high-energy excitations much above the Fermi level (the chemical potential for zero temperature)  become relevant and contribute to the plasmons while the effect of the flat band at zero energy is less noticeable . 

\medskip
\par
For $k_0=0$, we have obtained a plasmon branch relatively close to the diagonal $\omega=v_F q$,  which becomes Landau damped at the level of the band gap $\hbar \omega = \Delta_0$ and connects to the single-particle transitions from the flat band to the free states right about the Fermi energy in the conduction band. Instead, the energy and shape of the plasmon branch depend on the relative dielectric constant  $\alpha_r$, as can be verified by Eq.\,\eqref{eps01}. If $\alpha_r$ is large, the plasmon is found at higher frequencies further away from the diagonal, which displays a more nonlinear dependence. For a regular Lieb lattice with the Fermi level close to the flat band and the lowest point of the conduction band, we have $E_F = \Delta_0$ (under critical doping), and do not find a plasmon branch since both real and imaginary parts of the dielectric function never cross the zero line. This simply indicates that the plasmon mode does not exist in this case, which agrees well with our previous prediction in Ref.\,[\onlinecite{zhemchuzhna2024polarizability}].On the other hand, we find a well-formed plasma branch for the case having a higher doping level for $E_F = 2.0\, \Delta_0$. For this situation, we find a Landau damped plasmon mode at the level of $\hbar \omega = \Delta_0$. Compared to the previously considered case with $k_0  = 0$,  these plasmon branches are located at higher energies for the same values of $\alpha_r$, which is possibly due to the increased electron doping of the Lieb lattice. Consequently, we make the first important statement that one can find a well-formed plasma branch with a long lifetime over a wide range of energy and wave vector when the doping level of a Lieb lattice material is sufficiently high. This is one of the major results of our present work.

\medskip
\par
\section{Two Coulomb-coupled Lieb lattice layers}
\label{sec4}

An interesting issue for studying plasmon modes and other collective excitations in a quantum system is the dynamics of two Coulomb-coupled mono-layers of Lieb lattice. For this situation, two plasmon excitation modes in each layer will be coupled together by the Coulomb interaction between electrons in two different layers, which makes this system behave like two interacting quantum oscillators.  One of the first crucial results on such coupled plasmons in the two interacting layers of the 2D electron gas\,\cite{sarma1981collective} revealed that the spectrum of a lower-frequency branch exhibits a linear dependence. 

\medskip
\par
The plasmon modes of such Coulomb-coupled  system embedded within a uniform background dielectric constant are determined by the following determinant equation

\begin{equation}
\label{deteq01}
\text{Det} \left\{
\begin{array}{cc}
1 - V_i(q) \, \Pi_{1}(q,\omega)  &  V_c(q) \, \Pi_{2}(q,\omega) \\
V_c(q) \, \Pi_{1}(q,\omega) & 1 - V_i(q) \, \Pi_{2}(q,\omega) 
\end{array}
\right\} = 0 \ , 
\end{equation}
which is equivalent to

\begin{equation}
\Big\{ 
1 - V_i(q) \, \Pi_{1}(q,\omega) 
\Big\}\,
\Big\{ 
1 - V_i(q) \, \Pi_{2}(q,\omega) 
\Big\} - V_c^2(q) \, \Pi_{1}(q,\omega)  \, \Pi_{2}(q,\omega) = 0 \ . 
\end{equation}
Here, the intra-layer Coulomb potentials $ V_i(q)  = 2\alpha_r \, e^2/\pi q$, while the inter-layer one takes the form of 

\begin{equation}
 V_c(q)  = \frac{2 \pi  e^2}{q} \, \tet{e}^{- q d_{12}} = V_i(q) \, \tet{e}^{- q d_{12}}\ ,
\end{equation}
where $d_{12}$ is the separation between two coupled layers. 
\medskip 

From Eq.\,\eqref{deteq01}, it becomes clear that two individual layers equally participate in forming two hybridized plasma branches since this equation is symmetric with respect to switching the layer indexes. Here, two layers could be either identical or different by choosing electron doping or the Fermi  level. Furthermore, as the separation $d_{12}$ between two layer is increased, the inter-layer interactive term, {\em i.e.\/} the product of two polarization functions with an exponential factor, will decrease exponentially. As a result, the effective dielectric function of these two layers reduces to the product of two individual dielectric functions for two layers, {\em e.g.\/} by comparing Eq.\,\eqref{deteq01} with Eq.\,\eqref{eps01}.

\medskip 
\par
Our numerical results for plasmons in two Coulomb-coupled layers embedded in a uniform background dielectric constant for the Lieb lattice are displayed in Fig.\,\ref{FIG:5}. Here, we consider three different cases, including: two layers with $E_F=E_F^{(0)}$ (under-critical doping level) and $k_0=0$ in panels $(a)$ and $(d)$; and two layers with $E_F=E_F^{(0)}$ and $k_0=k_F^{(0)}$ in panels $(b)$ and $(e)$, as well as two layers with $E_F=2.0\,E_F^{(0)}$ $k_0=k_F^{(0)}$ in panels $(c)$ and $(f)$. From this figure, we do not observe any plasmon excitations when $E_F=\Delta_0$, similar to the case of a single Lieb-lattice layer. The effective dielectric function for two layers, as given by Eq.\,\eqref{deteq01} and shown on the lower panel $(b)$, never crosses zero (only asymptotically approaches it). Therefore, we will not see either damped or undamped plasmon modes, same as the case of an isolated layer. For two layers with different  doping levels but only one of them supports a plasmon mode,  we still find only one isolated plasmon branch. However, this branch has been divided into two parts, and only the long-wavelength part ($q \rightarrow 0$) demonstrates a nearly-linear spectrum. Finally, for two identical layers with the same doping $E_F=2\,\Delta_0$, we find rather traditional picture of two plasmon branches, {\em i.e.\/}, a linear acoustic low-frequency branch and another $\backsim \sqrt{q}$ optical branch in connection with higher frequencies. 

\medskip
\par

\section{Two-dimensional layer interacting with a surface plasmon}
\label{sec5}

A significant modification of the plasmon dispersions, energies and particle-hole modes could be achieved by bringing our 2D layer of Lieb lattice in electrostatic contact with a surface-plasmon mode in a semi-infinite metallic material. These Coulomb-coupled hetero-structures, which include a 2D layer ({\em e.g.\/}, a graphene layer) and a semi-infinite conductor, are sometimes referred to as open systems.\,\cite{iurov2017controlling} The most important and distinguished feature in these semi-infinite systems is the screened electrostatic interaction between electrons in both the 2D layer and the conducting substrate. Physically, these two types of plasmonic excitations appear in a way similar to two coupled quantum oscillators.
\medskip
\par

Previously, it was shown that, if a heterostructure cpnsists of one gapped graphene layer, its non-local collective modes become free from Landau damping within a much broader region in frequency-momentum space.\,\cite{gumbs2015nonlocal} In this sense, our current scheme for a Coulomb-coupled system, involving both a Lieb lattice layer and a surface plasmon, is an attempt to search for extended plasmon modes with a much longer lifetime, which is absent for a single-layer system. 

\medskip
\par
For a coupled system involving two plasmons within a layer and another localized at a surface, the previous single-layer dielectric function $\epsilon(q,\omega \, \vert \, \Delta_0)$ in Eq.\,\eqref{eps01} should be replaced by the so-called “dispersion factor”

\begin{equation}
\label{dispf01}
S_c(q,\omega)  = 1 - \frac{2 \pi \alpha}{q} \, \Pi^{(0)}(q,\omega) \left\{
1 + \tet{exp}(-2 q \,d_{12}) \, \frac{\Omega_p^2}{2 \omega^2 - \Omega_p^2}
\right\} \,  , 
\end{equation}
where $d_{12}$  is the distance between the 2D layer and the surface and  $\Omega_p = \sqrt{n_0 e^2/(\epsilon_0 \epsilon_r \,m^*)}$ is the bulk plasma frequency of a metal, which could be varied by changing the electron density $n_0$ over a substantial range between the terahertz and mid-infrared frequencies. The dielectric function of the bulk metal is $\epsilon_B(\omega) = 1 - \Omega_p^2/\omega^2$. The surface-plasmon resonance occurs at $\Omega_p/\sqrt{1+\epsilon_d}$ which is approximated as $\Omega_p/\sqrt{2}$ for a substrate dielectric constant $\epsilon_d=1$.  

\medskip
\par
A linear plasmon branch was previously observed in an open system including gapped graphene, which was later confirmed experimentally. For a larger separation between a layer and the surface, the effect due to a surface will be reduced dramatically, and the resulting plasmon excitations would nearly replicate the plasmon of an isolated two-dimensional layer. On the other hand, a dark-blue area connected to the pole $\Omega =\sqrt{2}\,\Omega_p$ in Eq.\,\eqref{dispf01} always shows up in all our density plots, which is not an indication of a plasmon. 

\medskip
\par
Our numerical results for an open system, which includes a 2D layer of the Lieb lattice interacting with a semi-infinite conducting material, is presented in Figs.\,\ref{FIG:6} and \ref{FIG:7}. We first consider the case with  doping $E_F=\Delta_0$ and, surprisingly, we observe a well-formed plasmon with its frequency asymptotically approaching $\Omega_p/\sqrt{2}$ for wave vectors $q \geq k_F$. These plasmon modes are strongly Landau damped since they fall into particle-hole mode regions, but the real part of the dispersion factor crosses the zero level and splits off, as can be verified from the right panels of Fig.\,\ref{FIG:6}. This observation comes in a stark contrast with earlier investigated plasmons in a free-standing layer, as displayed in Figs.\,\eqref{FIG:4}$(e)$ and $(f)$. Therefore, we conclude that even though a plasmon could not exist in an isolated layer because of a nonzero dielectric function, the new interaction between this layer and surface plasmon in another semi-infinite metal can support a new damped plasmon. This is one of the most crucial results presented in this paper.

\medskip
\par
For the case with higher doping level $E_F=2.0\,\Delta_0$, we find a significant hybridization between two different plasmon branches, which could be defined as low-frequency acoustic and high-frequency optic branches, respectively. Here, the low-energy acoustic plasmon mode resembles the plasmon in an isolated layer of Lieb lattice. Particularly, when the effect of Coulomb coupling becomes strong for the layer separation $d_{12} = 1/2 k_F^{(0)}$, the acoustic plasmon branch is found staying at lower frequencies and closer to the diagonal $\omega = v_F q$, which appears more linearly compared to the acoustic plasmon of a single layer. The other optical plasmon branch acquires much higher frequencies, approaching  $\Omega_p/\sqrt{2}$, and is observed at much larger wave vectors where the acoustic-plasmon branch is fully damped. 

\medskip
\par

\section{Comparing the plasmon dispersions of the  Lieb and dice lattices }
\label{sec6}

The major goal of current study is to compare the dynamical polarizability, and the plasmon dispersions of two lattices with a finite bandgap and a flat band having different band structures. As part of this endeavor,  we consider the plasmon branches and related quantities for a dice lattice (symmetric flat band) as well as the Lieb lattice (non-symmetric elevated flat band) under similar conditions. Specifically, we investigate the case with electron doping $E_F = 1.2\,\Delta_0$  so that some intraband electron transitions are still permissible.  

\medskip
\par
\begin{figure} 
\centering
\includegraphics[width=0.8\textwidth]{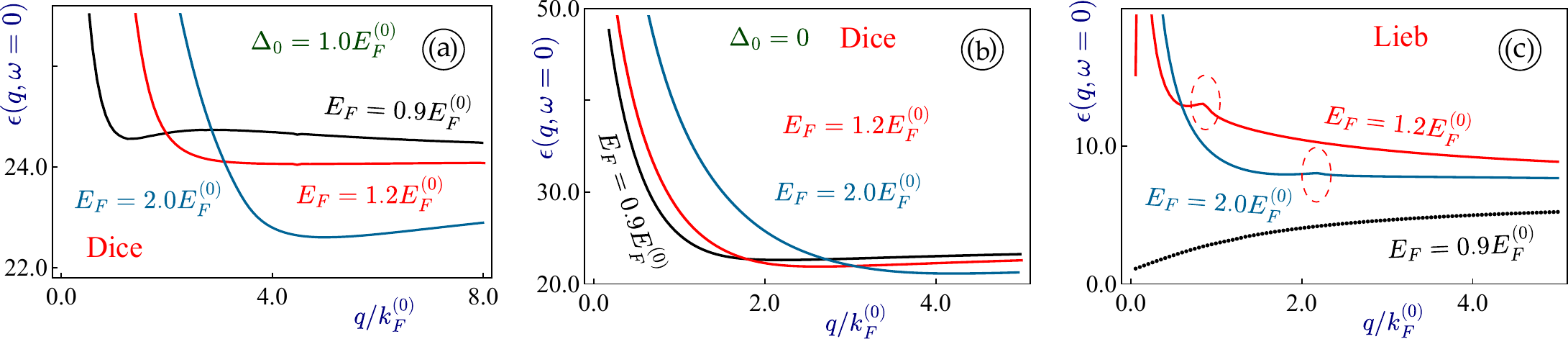}
\caption{(Color online) Static dielectric function $\epsilon(q,\omega=0 \, \vert \, \Delta_0)$ for a dice lattice as a function of wave vector $q$. Panels $(a)$ and $(b)$ demonstrate the static dielectric function for a finite- and zero-gap dice lattices, while plot $(c)$ - for the Lieb lattice. For all panels, each curve corresponds to a fixed Fermi energy (doping level) $E_F= 0.9 E_F^{(0)}$, $1.2E_F^{(0)}$ and $2.0 E_F^{(0)}$, according to the provided labels.}
\label{FIG:11}
\end{figure}

As we discussed above, the polarization function is determined by all possible electron transitions from occupied to unoccupied states (and backwards) between the valence and conduction bands.  In this case, the possibility for intraband transitions (within the conduction band through the forbidden energy level) becomes minimal. For this case, we find that the characteristics of the particle-hole modes are very different for the Lieb lattice and a dice lattice, {\em e.g.\/}, their difference due to the bandgap (comparing panels $(b)$ wand (c) of Fig.\,\ref{FIG:11}) is much less noticeable.  First and foremost,   we do not find any distinguished region satisfying $\text{Im} \left[ \Pi^{(0)}(q,\omega \, \vert \, \Delta_0) \right]  = 0$. For the Lieb lattice, this means that the plasmon becomes Landau damped for all considered frequencies and wave vectors. Meanwhile, we also find an extended region for very strong plasma damping below the diagonal $\omega=v_F q$, which results from intraband single-particle excitations ({\em i.e.}, the energy difference for transition between initial and final states of electrons within the same conduction band, (see Ref.\,[\onlinecite{ross2025dynamical}] for more detailed explanations). In contrast, zero-gap and finite-gap dice lattices exhibit a clear region for $\text{Im} \left[ \Pi^{(0)}(q,\omega \, \vert \, \Delta_0) \right]  =  0$ below the energy level $\hbar\omega  = \Delta_0$ and above the diagonal $\omega=v_F q$. 

\medskip
\par

The real part of the polarization function not only determines the plasmon spectrum of a considered material, but it is also very different for these two lattices. Meanwhile, some monotonic  and positive features are observed for the real part of the dynamical polarization function above the diagonal $\omega=v_F q$. This leads us to conclude that the dielectric function defined in Eq.\,\eqref{eps01} will never reach zero, and then, a plasmon mode cannot be achieved in this case, either Landau damped or not. However, such a situation will return to normal once the doping is increased. 

\medskip
\par

As a matter of fact, we do not detect clearly defined plasmon modes in these two cases, and the dielectric function in Fig.\,\ref{FIG:9} never reaches zero. However, our results still reveal essential and very interesting differences in  the plasmon excitations. By comparing panels $(a)$ and $(b)$ of Fig.\,\ref{FIG:9}, we note that the formation of a plasmon mode for the Lieb lattice is found only for very small values of wave vectors and frequencies, while the most noticeable part of dielectric function for the dice lattice stays along the diagonal next to the Fermi level.

\medskip
\par

Here, we also present the dependence at constant frequency for the real part of the polarization function in Fig.\,\ref{FIG:10}. As depicted in Fig.\,\ref{FIG:10}, there exist multiple peaks and non-monotonic features, thereby indicating the possibility for the presence of a plasmon mode in a dice-lattice structure. For high frequencies, there exist multiple peaks and non-monotonic features, likely leading to the existence of a plasmon mode. Additionally, the dependence on chemical potential (Fermi-level doping) becomes noticeable only for low frequencies and small wave vectors. 

\medskip
\par

\section{Static screening of a dilute distribution of charged impurities}
\label{sec7}

Let us consider a dilute distribution of non-magnetic charged impurities where the average separation between these charges is larger than the range of interaction between them. In this case, the  screened potential due to a dielectric medium in the vicinity of  a point charge $Q_0$ is proportional to the Fourier transformation of the screened electron-electron interaction, written as 

\begin{equation}
\Phi(r) = \frac{Q_0}{\alpha_r \epsilon_0} \int \frac{d^2\mbox{\boldmath${q}$}}{(2 \pi)^2} \, \frac{V_C(q)}{\epsilon(q, \omega \, \vert \, \Delta_0)} \, \tet{exp} \left(i\mbox{\boldmath${q}$}\cdot\mbox{\boldmath${r}$}\right) =  \frac{Q_0}{\alpha_r\epsilon_0} \int_0^\infty d q \,  \frac{J_0(q r)}{\epsilon(q, \omega \, \vert \, \Delta_0)} \ , 
\label{screen}
\end{equation}
where $J_0(q r)$ is the zero-order Bessel function. 
Here, the dominant $r$-dependence for the screened potential of a charged impurity in Eq.\,\eqref{screen} is determined by the Thomas-Fermi decay $\backsim 1/r^3$ for most known 2D materials.  

\medskip
\par
 
It is well known\,\cite{lighthill1958introduction, gamayun2011dynamical} that the asymptotic dependence and spatial feature of a screened potential is determined by all non-analytical points (discontinuity in the function or its derivative) of a dielectric function as well as the denominator of  the equation, and this results in well-known  Friedel oscillations. The key difference between graphene and a dice lattice is the smooth behavior of the polarization function (also called Lindhard function) in the static limit $\omega\rightarrow 0$,  as well as in the dielectric function. This implies that, for a dice lattice, its first derivative should be continuous and its first non-analytical point will be found in third order. Therefore, the singularity of any derivative of the integrand in a potential expression can predetermine the asymptotical dependence of the screened potential associated with a charged impurity. Besides, it is also known that the Friedel oscillations are  proportional to $\backsim \cos(2 k_Fr)/r^3$ at T=0K. Therefore, for a  dice lattice, we expect its potential proportional to $\backsim \cos(2 k_Fr)/r^4$. 

\medskip
\par

We show numerically computed results of  the dielectric function $\epsilon(q,\omega=0 \, \vert \, \Delta_0)$ for a dice and the Lieb lattice in Fig. \,\ref{FIG:11}. Analytical expressions have been derived for the static dielectric function for graphene, as well as for the zero gap dice lattice, as shown in panel $(b)$ of Fig.\,\ref{FIG:11}. It consists of two terms: one persists for all values of wave vector $q$, while the other is only relevant for $q > 2 k_F$.  The key difference between a dice lattice and the Lieb lattice is that we observe a smooth $q$ dependence for both static polarization function and the dialectic function in a dice lattice. However, for the Lieb lattice,  we find the points of  discontinuity in the first derivative of its integrand (breaking points shown by a red dashed circle in panel ($c$) of Fig.\,\ref{FIG:11}). Therefor, we conclude that the static screening and the Friedel oscillations are very different in the Lieb lattice in comparison with that of dice lattice which is similar to that of graphene.  

\medskip
\par

\section{Summary and Concluding remarks}
\label{sec8}

In this work,  our key goal has been to investigate the plasmon modes, their dispersions, and Landau damping for various conditions and configurations of free-standing  Lieb lattice structures. In some sense,  this could be viewed as a continuation of our previous work\,\cite{zhemchuzhna2024polarizability} on the Lieb lattices, in which we could only find a plasmon mode within a limited momentum-energy range. 

\medskip
\par
We have provided a detailed comparison between the plasmonic properties of the Lieb lattice and a dice lattice. Both of these materials acquire a finite bandgap and a flat band in their low-energy band structure. A significant difference between them is that  the flat band is located in an elevated position and intersects the conduction band at its lowest point for a Lieb lattice. In contrast, the band structure of a gapped dice lattice is symmetric, so that its flat band is located right in the middle of a gap between the valence and conduction bands. Our results are based on detailed numerical computations of the dynamical polarization function and the dielectric function under the random-phase approximation (taking into account all possible electron transitions from occupied to empty states and back) as its most important building block. We have found that for under-critical doping (below or approaching the flat band) of the Lieb lattice, the computed dynamical polarization function is nearly monotonic and always negative, which excludes the presence of a plasmon mode.

\medskip
\par

The first and  foremost idea that could help to find a well-formed plasmon excitation is adjusting the doping level or Fermi energy at zero temperature. By raising the doping level, the number of allowed intraband transitions of electrons by going through the Fermi level of conduction band and with a small momentum transfer (long-wavelength limit) is greatly increased, which is able to facilitate the existence of a plasmon mode having a long lifetime. Apart from that, the major difference between the Lieb lattice and a dice lattice, for which we always observed a plasmon, is the location of a flat band. However, the difference due to the location of a flat band becomes less relevant once the doping level becomes significantly high. As it appears,  the idea of increasing electron doping has worked out very well, and we do find a plasmon mode within a high-frequency range.  We have explored various other ways to obtain a well-formed plasmon apart from increasing the level of electron doping, including considering a double layer and a mono-layer of the Lieb lattice, as well as Coulomb-coupling with a semi-infinite conductor. In both cases, the two plasmon modes interact with each other, resemble to two coupled quantum oscillators. We have also found that even in the situation where a plasmon could not be found or observed in an isolated layer because the dielectric function is never equal to zero. However, by bringing the same layer under similar conditions into a Coulomb contact with a surface plasmon of a semi-infinite metal, this leads to the presence of a new damped plasmon mode.

\medskip
\par

We have provided a detailed comparison for the polarization functions, plasmon dispersions, and their Landau dampings for a dice and Lieb lattice under  similar conditions and with equivalent electron doping levels. We have further discovered and analyzed several crucial distinguishing features of the Lieb lattices, {\em e.g.\/}, the plasmon mode is formed at larger wave vectors and higher frequency along the diagonal $\omega = v_F q$, in contrast to a dice lattice in which the plasmon mode is found only in the long-wavelength limit $q \rightarrow 0$.  We have also considered the static screening and the screened potential of a charged impurity, which is a crucial part for calculating various transport properties, such as semi-classical Boltzmann conductivity. We have further found a crucial difference between a dice lattice and the Lieb lattice, namely, in the latter case, we have observed a discontinuity in the first derivative of both static polarization function and dielectric function. This implies that the physics nature and the features of Friedel oscillations for these two lattices are quite different, and the Lieb lattice is found much closer to a regular graphene than a pseudospin-1 dice lattice.  

\medskip
\par

In general, the collective excitations, such as plasmons, excitons and polaritons, are crucial for understanding the fundamental physical properties of materials. These include all recently discovered innovative ones, and we strongly believe that our presented work reveals a trove of novel and crucial knowledge about the fundamental electronic properties of these low-dimensional materials. We are also confident that our new results presented in this paper will be useful for multiple promising applications in electronic and plasmonic nano-devices. 

\medskip
\par

\begin{acknowledgements}
A.I. was supported by the funding received from TradB-56-75, PSC-CUNY Award \# 68386-00 56. G.G. gratefully acknowledges funding from the U.S. National Aeronautics and Space Administration (NASA) via the NASA-Hunter College Center for Advanced Energy Storage for Space under cooperative agreement 80NSSC24M0177.  D.H. would like to acknowledge the Air Force Office of Scientific Research (AFOSR). The views expressed are those of the authors and do not reflect the official
guidance or position of the United States Government, the Department of Defense, or the United States Air Force.
\end{acknowledgements}
\medskip

\bibliography{DLP}
\end{document}